\documentclass[final]{interact}

\usepackage{natbib}
\bibpunct[, ]{[}{]}{,}{n}{,}{,}% Citation support using natbib.sty

\usepackage{amsmath}
\usepackage{float}
\usepackage{mathtools}
\usepackage{epstopdf}% To incorporate .eps illustrations using PDFLaTeX, etc.
\usepackage[caption=false]{subfig}% Support for small, `sub' figures and tables
\usepackage{framed} % Framing content
\usepackage{multicol} % Multiple columns environment
\usepackage{nomencl} % Nomenclature package
\usepackage{xcolor}

\setcitestyle{authoryear,sort,round,citesep={; },aysep={}}

\renewcommand\bibfont{\fontsize{10}{12}\selectfont}% Bibliography support using natbib.sty
\makeatletter% @ becomes a letter
\def\NAT@def@citea{\def\@citea{\NAT@separator}}% Suppress spaces between citations using natbib.sty
\makeatother% @ becomes a symbol again

\theoremstyle{plain}% Theorem-like structures provided by amsthm.sty

\theoremstyle{definition}

\theoremstyle{remark}
\newtheorem{remark}{Remark}

\begin{document}

\articletype{Technical Paper}% Specify the article type or omit as appropriate

\title{Relating unsaturated electrical and hydraulic conductivity of cement-based materials}

\author{
\name{D.~J. Smyl\textsuperscript{a}\thanks{CONTACT: D.~J. Smyl. Email: danny.smyl@aalto.fi} }
\affil{\textsuperscript{a}Department of Mechanical Engineering, Aalto University, Espoo, Finland}
}

\maketitle

\begin{abstract}
Unsaturated hydraulic ($K$) and electrical ($\sigma_b$) conductivity are often considered durability indicators of cement-based materials.
However, $K$ is difficult to measure experimentally.
This is due to the large pressure requirements at low degrees of saturation resulting from the fine pore-size distribution of cement-based materials.
As a result, the commonly-used analytical models, requiring calibration of $K$ from experimental data, are often inaccurate at low degrees of saturation.
On the other hand, measuring $\sigma_b$ is rather straight forward.
Descriptions of the relationship between $\sigma_b$ and $K$ may therefore be particularly valuable when $K$ is required.
In this work, we use experimental data from previous works to determine the feasibility of models employing a van Genuchten-Mualem based framework to predict $K$ and $\sigma_b$ -- {expressions for diffusivity $D$ are also provided.}
We then develop analytical expressions relating $K$ and $\sigma_b$ using these models.
It is then shown that $K = K(\sigma_b)$ and $\sigma_b = \sigma_b(K)$ may be determined when either parameter is fully described.
Lastly, we propose a simplified model and discuss the roles of pore-size distribution, saturation, pore connectivity and tortuosity in characterizing the relationship between $K$ and $\sigma_b$.

\end{abstract}

\begin{keywords}
Cement-based materials, transport properties, unsaturated moisture transport
\end{keywords}

\pagebreak
\tableofcontents

\section*{Nomenclature}
\begin{tabular}{l l} 
{$A$}&{Archie's coefficient, $(\mathrm{-})$}\\
{$D$}&{unsaturated water diffusivity, $(\mathrm{mm^2/hr})$}\\
{$D_0$}&{limiting diffusivity, $(\mathrm{mm^2/hr})$}\\
{$g$}&{conversion factor, $(\mathrm{mm})$}\\
{$h$}&{capillary pressure head, $(\mathrm{mm})$}\\
{$i$}&{ingress, $(\mathrm{mm^3/mm^2})$}\\
{$I$}&{tortuosity/pore-connectivity parameter, $(\mathrm{-})$}\\
{$K$}&{unsaturated hydraulic conductivity, $(\mathrm{mm/hr})$}\\
{$K_r$}&{relative hydraulic conductivity, $(\mathrm{-})$}\\
{$K_s$}&{saturated hydraulic conductivity, $(\mathrm{mm/hr})$}\\
{$m$}&{van Genuchten parameter, $(\mathrm{-})$}\\
{$n$}&{van Genuchten parameter, $(\mathrm{-})$}\\
{$q$}&{van Genuchten parameter, $(\mathrm{-})$}\\
{$RH$}&{relative humidity, $(\mathrm{-})$}\\
{$w$}&{fitting parameter for $\sigma_p$, $(\mathrm{-})$}\\
{$w/c$}&{water-to-cement ratio, $(\mathrm{-})$}\\
{$s$}&{reduced sorptivity, $(\mathrm{mm/\sqrt{hr}})$}\\
{$S$}&{sorptivity, $(\mathrm{mm/\sqrt{hr}})$}\\
{$z$}&{``critical exponent", $(\mathrm{-})$}\\
{$\alpha$}&{van Genuchten parameter, $(\mathrm{1/mm})$}\\
{$\beta$}&{shape parameter, $(\mathrm{-})$}\\
{$\gamma$}&{fitting parameter, $(\mathrm{-})$}\\
{$\sigma_b$}&{unsaturated bulk electrical conductivity, $(\mathrm{S/m})$}\\
{$\theta$}&{volumetric moisture content, $(\mathrm{mm^3/mm^3})$}\\
{$\theta_d$}&{dynamic moisture content, $(\mathrm{mm^3/mm^3})$}\\
{$\theta_s$}&{volumetric moisture content at saturation/open porosity, $(\mathrm{mm^3/mm^3})$}\\
{$\Theta$}&{degree of saturation, $(\mathrm{mm^3/mm^3})$}\\
{$\Theta_c$}&{``critical degree of saturation", $(\mathrm{mm^3/mm^3})$}\\
{$\Theta_{\delta^-}$}&{Infinitesimally lower degree of saturation, $(\mathrm{-})$}\\
{$\zeta$}&{pore connectivity, $(\mathrm{-})$}\\
{$\kappa$}&{fitting parameter, $(\mathrm{\frac{S\cdot hr}{mm^2}})$}\\
{$\lambda$}&{Brooks and Corey's fitting parameter, $(\mathrm{-})$}\\
{$\sigma_p$}&{electrical conductivity of the pore solution, $(\mathrm{S/m})$}\\
{$\sigma_r$}&{relative electrical conductivity, $(\mathrm{-})$}\\
{$\sigma_s$}&{saturated bulk electrical conductivity, $(\mathrm{-})$}\\
{$\Omega$}&{pore-size distribution, $(\mathrm{-})$}\\
{$\Upsilon_{d/a}$}&{desorption or absorption isotherm, $(\mathrm{-})$}\\
{$\phi$}&{porosity in Archie's model, $(\mathrm{-})$}\\
{$\theta_i$}&{initial moisture content, $(\mathrm{mm^3/mm^3})$}\\
{$\theta_d$}&{dynamic moisture content, $(\mathrm{mm^3/mm^3})$}\\
{$\theta_r$}&{residual moisture content, $(\mathrm{mm^3/mm^3})$}\\
{$\phi_c$}&{percolation threshold, $(\mathrm{-})$}\\
{$\tau$}&{tortuosity, $(\mathrm{-})$}\\
{$\rho_b$}&{resistivity, $(\mathrm{ohm \cdot m)}$}\\

\end{tabular}

%%%%%%%%%%%%INTRO
\section{Introduction}
\label{sec.introduction}

Electrical and hydraulic properties are used as durability indicators of cement-based materials \citep{Ghasemzadeh, Castro, HallHoff, Martys}.
However, often only the transport properties that are straight-forward to obtain experimentally are used as indicators.
These properties are commonly the sorptivity $S$, saturated hydraulic conductivity $K_s$, open porosity $\theta_s$, electrophorosis-related properties (rapid chloride penetration, RCP), and the desorption/adsorption isotherm ($\Upsilon_{d/a}(\Theta)$, where $0 \leq \Theta < 1.0$ is the degree of saturation).
While these parameters are related to unsaturated transport, none completely paints the picture of unsaturated transport in cement-based materials \citep{zhou2016indirect,Smyl, Scherer}.
A more descriptive unsaturated transport property is the unsaturated hydraulic conductivity $K$.
However, obtaining $K$ for cement-based materials using experimental methods is extremely difficult due to the large pressure requirements at low saturation due to the fine pore-size distribution.
Specifically, the challenge in directly measuring $K$ results from maintaining a steady level of unsaturation while simultaneously supplying water at the interface of inflow \citep{HallHoff}.
For this reason, analytical models predicting $K$ are often used, often the van Genuchten-Mualem model.
Due to the lack of corroborating data, however, the tortuosity/pore-connectivity parameter $I$ required for this model is commonly estimated using a default value ($I=$ 0.5).
As a result, current models may under predict $K$ by orders of magnitude at low $\Theta$ \citep{Poyet2011}.

Another important transport parameter is the unsaturated electrical conductivity $\sigma_b$.
Unlike $K$, $\sigma_b$ is simple to measure experimentally \citep{spragg2013factors,weiss2012using}.
Although, developing prediction models for $\sigma_b$ often requires knowledge of the pore solution conductivity, $\sigma_p$.
Obtaining $\sigma_p$ directly requires high-pressure techniques or chemical models prone to inaccuracies at low $\Theta$ \citep{weiss2012using,rajabipour2007electrical}.
Therefore, contemporary prediction models that only consider $\Theta$, $\sigma_p$, and porosity often diverge from experimental measurements \citep{li2016effect}.

Recently, there has been much research interest in understanding connections between experimentally-obtained transport properties ($S$, $\Upsilon_{d/a}(\Theta)$, $\theta_s$, and $RCP$) and more ``characteristic" measures of unsaturated transport properties ($K$ and $\sigma_b$) \citep{ghasemzadeh2016comparison,YF,Hallaji2015,Pour-Ghaz2011a,neithalath2006}.
However, due to the complexity in the relationships between these properties, very few analytical or empirical connections between these properties are available.
Indeed,  in the recent paper by \cite{li2016effect}, the authors stated that the relation between hydraulic properties and $\sigma_b$ is poorly understood, especially when considering the microporous structure of cement-based materials.
In particular, the collective roles of pore-size distribution ($\Omega$, obtained from $\Upsilon_{d/a}$), $I$, $\Theta$, and $\theta_s$ require further research to understand their roles in $K$ and $\sigma_b$.

We begin this paper by reviewing the theory and methods related to modeling $K$ and $\sigma_b$.
Following, we describe the roles of $\Omega$, $I$, $\Theta$, and $\theta_s$ in the van Genuchten-Mualem framework we use to model $K$ and $\sigma_b$.
To address the issue of lacking experimental data for the determination of $K$, we determine $K$ using $\Omega$ and $S$.
Similarly, we describe $\sigma_b$ using experimentally-measured parameters.
%We note that relation of $RCP$ to $K$ and $\sigma_b$ is ommited in this paper.
Next, we develop analytical expressions for $K=K(\sigma_b)$ and $\sigma_b = \sigma_b(K)$ and apply the models using experimental data.
Finally, we propose a simplified model and discuss the roles of $\Omega$, $\Theta$, pore connectivity, and tortuosity in characterizing the relationship between $K$ and $\sigma_b$.
The central aims of this paper are enumerated below.\\

\begin{enumerate}
%\item Describe the roles of $\Omega$, $I$, $\Theta$, and $\theta_s$ in modeling $K$ and $\sigma_b$.
\item Determine the feasibility of modeling $K$ and $\sigma_b$ from experimental data using the van Genuchten-Mualem models, considering\footnote{The related parameter, water Diffusivity ($D$), is also considered.}:
\begin{center}$ K = K(S,\Omega, I, \Theta, \theta_s)$\end{center}
\begin{center}$ \sigma_b= \sigma_b(\sigma_p, \Omega, I, \Theta, \theta_s)$\end{center}
\vspace{-2mm}

\item Develop an analytical model relating $K$ and $\sigma_b$:
\begin{center}$K = K(\sigma_b)$\end{center}
\begin{center}$\sigma_b = \sigma_b(K)$\end{center}
\vspace{-2mm}

\item Develop a simplified model to characterize the roles of $\Omega$, $\Theta$, pore connectivity, and tortuosity on $K$ and $\sigma_b$.

\end{enumerate}
\vspace{-8mm}

%%%%%%%%%%SECTION%%%%%%%%%%
\section{Background -- theory and modeling of {hydraulic and electrical conductivity}}
This section provides a brief description of the theory and modeling of electrical and hydraulic conductivity.
A central theme in the conduction of fluids and electric current in porous materials is the connectivity of the pores \citep{Ulm}.
Perhaps the most simple model to describe pore connectivity is percolation theory.
Percolation theory describes the connectivity of the pore system $\zeta(\phi,z)$ in terms the porosity $\phi$, phenomenologically this is written as 

\begin{equation}
\zeta(\phi,z) \propto (\phi - \phi_c)^z
\label{perc}
\end{equation}

\noindent where $\phi_c$ is the threshold porosity and $z > 1.0$ is a ``critical exponent" related to the connectivity of the pore system \citep{Hunt,Bentz2}.
Here, we distinguish $\phi$ from the open porosity $\theta_s$, as these parameters are sometimes taken as different measures \citep{HallHoff}.
Above a certain value of $\phi_c$, the pore system is considered connected and capable of conducting electricity and transporting fluids \citep{durner1994hydraulic}.
While there is some debate on the true value of $\phi_c$ for cement-based materials, it is often approximated as 0.20 which was originally determined by \cite{Powers}.
$z$ is often taken from mathematical modeling or experiment; for example, \cite{Hunt2} found $z = 1.88$ for porous material with percolated spheres and \citep{Ghasemzadeh} estimated $z=3.0$ for cement-based material with distributed damage.
In terms of unsaturated hydraulic conductivity, $K$ may be conceptualized as 

\begin{equation}
K \propto K_s(\Theta - \Theta_c)^z
\label{perc2}
\end{equation}

\noindent where $\Theta_c$ is the ``critical degree of saturation" for water percolationand $K_s$ is the  hydraulic conductivity at saturation ($\Theta = 1.0$).
$\Theta_c$ may be interpreted as the degree of saturation, below which, there is no continuous water connectivity in the pore system  to drive capillary conduction.
For single phase flow, $\Theta_c=0$ is a generally assumed.
Similarly, an expression for $\sigma_b$ can be related to the electrical conductivity $\sigma_s$ at saturation as 

\begin{equation}
\sigma_b \propto \sigma_s(\Theta - \Theta_c)^z.
\label{perc3}
\end{equation}

These models offer insight on the relation of $K$ and $\sigma_b$ to the degree of saturation.
Functional dependence of pore connectivity may also be qualitatively interpreted from percolation theory, which indicates that the pore system is less connected at lower degrees of saturation \citep{Ye,Clemo}.
In concept, both hydraulic and electrical conductivity increase exponentially as a function of saturation and pore connectivity.
This trend is generally true for unsaturated porous material, however the addition of metallic- or carbon-based fibers may alter the behavior  \citep{chen2004conductivity, chung2000cement}.
Here, we restrict ourselves to studying non-fiber reinforced cement-based materials.

Quantitative expression of hydraulic and electrical conductivity in terms of saturation and porosity usually requires experimentally-measured or calibrated parameters.
In estimating the electrical conductivity of cement-based materials, Archie's law-based models are popular due to their simplicity \citep{sant2011capillary}.
Archie's law relates $\sigma_b$ to the electrical conductivity of the pore solution $\sigma_p$ and $\phi$.
Archie's law is written as

\begin{equation}
\sigma_b =A  \sigma_p \phi^z.
\label{Archie}
\end{equation}

In Equation \ref{Archie}, $z$ is different than in the percolation model (Equation \ref{perc3}).
In general, $A$ and $z$ may be taken as a calibration parameters.
Archie's law is a powerful and simple tool to predict the bulk electrical conductivity of porous media.
However, Archie's law is often not sufficient when precise knowledge of electrical conductivity is required (for example, in quantitative applications of Electrical Impedance Tomography \citep{Smyl3}).
Indeed, \cite{li2016effect} recently found that electrical conductivity predictions using Archie's law showed large deviation from experimental data at high saturation ($\Theta > 0.70$) using conventional values for $z$.
The authors concluded that the electrical conduction of cement-based materials is far to complex to be precisely modeled by applying only Archie's law and that more advanced methods should be studied.
Such a suggestion may imply that other factors directly accounting for the effects of tortuosity and/or pore size-distribution may be required.

Models estimating electrical conductivity of cement-based materials, directly accounting for pore-size distribution and tortuosity, are scarce.
However, researchers studying geophysical applications have developed models including these factors.
One of the first models including tortuosity and pore-size distribution was proposed by  \cite{mualem1991theoretical}.
Mualem and Friedman's model incorporated the \cite{brooks1964hydraulic} description of the water retention curve (yielding the saturation function $F(\lambda)$) and Mualem's ``capillary tube" hydraulic model.
Mualem and Friedman's model is written as

\begin{equation}
\sigma_b = \sigma_s F(\lambda) \frac{ \theta^{I + 2} }{\theta_s}, ~F(\lambda) = \frac{1+\frac{2}{\lambda}}{(1+\frac{1}{\lambda})^2}
\label{MualemSigma0}
\end{equation}

\noindent where $\lambda > 0$ is the Brooks-Corey water-retention curve fitting parameter, $\theta_s~\mathrm{(mm^3/mm^3)}$ is the volumetric moisture content at saturation, $\theta= \theta_s \Theta~\mathrm{(mm^3/mm^3)}$ is the unsaturated volumetric moisture content  (neglecting residual water content \citep{Pour-GhazHaifa}), and $I(-)$ is the tortuosity and pore conductivity parameter.
While the model is commonly used in geophysical applications, it has been noted to misrepresent the electrical conductivity of porous materials at high saturation and $F(\lambda)$ asymptotically approaches 1.0 for large values of $\lambda$ \citep{amente2000estimation}.

Comparatively speaking, models for the hydraulic conductivity of cement-based materials are more developed than models for electrical properties \citep{AkhavanCCR,HallHoff,Grassl, Wadso,Baroghel-Boundy2007a, Daian}.
Nearly all contemporary models of moisture retention and unsaturated hydraulic conductivity include, at a minimum, pore-size distribution and moisture saturation.
To date, one of the most popular models for unsaturated hydraulic conductivity is the van Genuchten-Mualem model \citep{VG1980}.
To describe $K$ of porous materials, van Genuchten proposed the expression:

\begin{equation}
K = K_s K_r = K_s \Theta^I (1 - (1 - \Theta^{1/m})^m)^2
\label{MualemSigma}
\end{equation}

\noindent where $(0 \leq K_r < 1.0)$ is the relative hydraulic conductivity and $0<m<1.0$ is closely related to pore-size distribution $\Omega$ of the material.
{It can, in fact, be shown through modulating the value of $m$, that $m$ is a rough measure of $\Omega$ since $m$ controls the bandwidth of the pore-size distribution approximated by $\mathrm{abs}( \frac{\partial \Theta}{\partial h})$ \citep{durner1994hydraulic}, where $h~\mathrm{(mm)}$ is the capillary suction.}

The model proposed by van Genuchten for $K$ is similar to that proposed by Mualem, using Brooks and Corey's moisture-retention model, however van Genuchten's model {for moisture retention} is more realistic near saturation. 
{The moisture retention model proposed by van Genuchten is given by}

{
\begin{equation}
\Theta = \frac{1}{(1+(\alpha h)^n)^m}, ~ n = \frac{1}{1-m}
\label{VG}
\end{equation}
}

\noindent { where $n(-) > 1.0$ and $\alpha~\mathrm{(mm^{-1})} > 0$ are fitting parameters.}
While the sensitivity of Equation \ref{MualemSigma} to parameters $m$ and $\alpha$ is well understood, until only recently, {the influence of the tortuosity and pore-connectivity parameter} $I$  was not well understood in cement-based materials \citep{Poyet2015, Poyet2011}.
Often, a default value of $I = 0.5$ (proposed my Mualem as a good fit for 45 undisturbed soils) is assumed \citep{Scherer}.
\cite{Poyet2011} showed that the assumption of $I = 0.5$ led to poor estimation unsaturated hydraulic conductivity.
In general, the researchers found that negative values of $I$ often resulted in the best modeling of experimental data.
Poyet et al. developed an expression for the tortuosity and pore connectivity parameter as a function of $m$, (i.e. $I = I(m)$).
However, they noted that more experimental data is required to broaden the applicability of their expression. 
{The presence of $I$ in hydraulic modeling is therefore a central source of uncertainty.
We will show in this work that, by relating $K$ to transport parameters such as $S$ and $\sigma_b$, direct dependence on $I$ may be eliminated.}
In the following section, {we begin by using the van Genuchten model to construct expressions for hydraulic conductivity and diffusivity as a function of $S$ and independent of $I$.}

\section{Modeling approach for {hydraulic conductivity and diffusivity} using sorptivity data}
\subsection{General - review of transport parameters $K$ and $D$}

In this section we show that $S$ and $\Omega$ may be used to determine $K$ from the unsaturated capillary water diffusivity $D$.
First, we provide a brief description of $D$ and $K$.
$D$ and $K$ are related unsaturated hydraulic properties that are part of the description of capillary flow in porous media \citep{celia1990general}.
The following one-dimensional partial differential equations describe capillary flow as a function of $K$:

\begin{equation}
\frac{\partial \theta}{\partial t} = \nabla \cdot (K \nabla h)
\label{Kpartial}
\end{equation}

or as a function of $D$:

\begin{equation}
\frac{\partial \theta}{\partial t} = \nabla \cdot (D \nabla \theta)
\label{Dpartial}
\end{equation}

where

\begin{equation}
D = K\frac{dh}{d\theta}.
\label{KD}
\end{equation}

In Equation \ref{KD}, the function $\frac{dh}{d\theta}$ is determined from the moisture retention curve (see Equation \ref{VG}) and using $\Theta = \frac{\theta}{\theta_s}$.
$\frac{dh}{d\theta}$ is a highly nonlinear function that has hysteresis, depending on the drying and rewetting history of the material \citep{wu2017T}.
Therefore, modeling $D$ or $K$ using a single function of $\frac{dh}{d\theta}$ should be done with some caution.
It should be noted here that determining $\frac{dh}{d\theta}$ (or $\theta(h)$) experimentally is rather straight forward \citep{ghasemzadeh2016comparison,Ghasemzadeh}, however determining $D$ or $K$ for cement-based materials using direct methods is very difficult \citep{bao2017empirical}.

%%%%%%SUBSECTION%%%%%%%%%
\subsection{Determining $D$ and $K$ using the sorptivity test}
\label{DFKG}

The sorptivity test measures the vertical ingress, $i~\mathrm{(mm^3/mm^2)}$, of water from the bottom surface of a cylindrical specimen as a function of time.
As a result, the sorptivity $S = \frac{i}{\sqrt{t}}$ may be directly computed.
In turn, $S$ may be related to $D$ using the approximation \citep{leech2003unsaturated}

\begin{equation}
s=\frac{S}{\theta_d - \theta_i} \approx \Big(\int_0^1 (1+ \Theta)D d\Theta \Big)^{1/2}
\label{S}
\end{equation}

\noindent where $s~\mathrm{(mm/\sqrt{t})}$ is the reduced sorptivity, $\theta_d~\mathrm{(mm^3/mm^3)}$ is the dynamic moisture content at saturation (which may be less that $\theta_s$ due to air trapping) and $\theta_i$ is the initial moisture content.
$\theta_d$ may be approximated using $\theta_d = \frac{iA}{V_s}$, where $A$ is the specimen's cross-sectional area and $V_s$ is the specimen volume \citep{Smyl}.

Using Equation \ref{S} allows for direct evaluation of the limiting magnitude term, $D_0$, in the well-known \citep{HallHoff,lockington1999} diffusivity power function given by $D = D_0\Theta^\beta$.
Using Equation \ref{S}, we may estimate $D_0$ using

\begin{equation}
D_0 = s^2\frac{(1+\beta)(2+\beta)}{(3+2\beta)}
\label{D0}
\end{equation}

\noindent where $\beta$ is a shape term generally considered to be bounded between 4-6 for cement-based materials \citep{leech2003unsaturated}.
Using the definition of $D$ and $K$, we may now relate these hydraulic properties using the equation\footnote{In order to determine $\frac{dh}{d\theta}$, van Genuchten assumed: $m +1/n -1 =0$.} proposed by van Genuchten \citep{VG1980} (assuming $\theta_r = 0$)

\begin{equation}
D =  K\frac{dh}{d\theta} = \frac{(1-m)K_s}{\alpha m \theta_s} \Theta^{I-1/m} \Big[(1-\Theta^{1/m})^{-m} + (1-\Theta^{1/m})^m \Big]^2.
\label{DK}
\end{equation}

By substituting $K_s = K/K_r$ ($K_r$ is in the form of Equation \ref{MualemSigma}) and rearranging, we eliminate dependence on $I$, yielding

\begin{equation}
D =  K \Bigg[\frac{\Theta^{-1/m}(m-1) \Big( (1-\Theta^{1/m})^m + \frac{1}{(1-\Theta^{1/m})^m}-2 \Big)}{\alpha m \theta_s \Big( (1-\Theta^{1/m})-1 \Big)} \Bigg].
\label{DKF}
\end{equation}

For simplicity, we write $D = Kg$ or $K = Dg^{-1}$, where $g$ is the expression within the brackets of Equation \ref{DKF}.
We note that Equation \ref{DKF} is undefined at $\Theta = 1.0$, and is therefore only valid in unsaturated conditions.
If we describe $D$ using the power equation, we may then write

\begin{equation}
K = D_0 \Theta^{\beta} g^{-1}.
\label{DKFP}
\end{equation}

Equation \ref{DKFP} provides a closed-form expression for $K$ when the moisture retention curve and sorptivity are known.
$K$ may also be determined from $K = Dg^{-1}$ when $D$ is determined experimentally, for example by long-term electromagnetic-based tomography or radiography results \citep{carmeliet2004D}.
On the other hand, when only $D$ is required, Equation \ref{DK} may be used with knowledge of $K_s$ and the moisture retention curve.
It is important to note that Equations \ref{DKF} and \ref{DKFP} tends toward zero near saturation, with the decreasing rate largely being controlled primarily by $m$ and $\beta$.
For this reason, the constraint $K(\Theta) \geq K(\Theta_{\delta^-})$ or $D(\Theta) \geq D(\Theta_{\delta^-})$ should be imposed, where $\Theta_{\delta^-}$ is an infinitesimally lower degree of saturation than the current degree of saturation.
For cement-based materials examined herein the tendency towards zero occurs at high degrees of saturation (e.g $\Theta  \gtrapprox  0.95$).

%%%%%SUBSUBSECTION%%%%%%
\subsection{Application}
\label{Dam1}
In this section, we use data from \citep{Smyl} (see Table \ref{T}) to estimate $K$ by first computing $D$.
The materials selected for this application were chosen (i) to test the robustness of the method and (ii) since the data includes all the required hydraulic parameters.
Note that in Table \ref{T}, the indicators ``M" and ``C" denote mortar and concrete, while the number after the indicator denotes the degree of freeze-thaw damage.
Further details of damage quantification and experimental freeze-thaw procedures may be found in \citep{ghasemzadeh2016comparison,Ghasemzadeh}.

\begin{table}[h]
\fontsize{8.5}{8.5}\selectfont
\renewcommand{\arraystretch}{1.35}
\caption{Hydraulic and electrical parameters for damaged mortar and concrete taken from \citep{Smyl}} 
\centering 
\begin{tabular}{l | c c c c | c c c c} 
\hline\hline 
Parameter & M0 & M18 & M30 & M48 & C0 & C21 & C29 & C47 \\ 
\hline 
\begin{math}{S} \end{math}$[\frac{mm}{\sqrt{hr}}]$ & 0.008 &$0.03$&$0.031$ & $0.044$ & 0.003 & $0.018$  & $0.033$ & $0.05$  \\
\begin{math}{\alpha} \end{math}$[\frac{1}{mm}]$ & $0.012$ & $0.018$ & $0.038$ & $0.017$ & $0.013$ & $0.046$ & $0.046$ & $0.043$ \\
\begin{math}{n}[-] \end{math} & 1.8& 1.8& 1.5 & 1.8 & 2.3& 2.5 & 2.5 & 2.3\\
\begin{math}{\theta_i} \end{math}$[\frac{ mm^3}{mm^3}]$ & 0.03& 0.010 & 0.020 & 0.010 & 0.03 & 0.040 & 0.030 & 0.030 \\
\begin{math}{\theta_{s}} \end{math}$[\frac{ mm^3}{mm^3}]$ & 0.14 & 0.156 & 0.176 & 0.200 & 0.15 & 0.162 & 0.170 & 0.192 \\
\begin{math}{K_{s}} \end{math}$[\frac{mm}{hr}]$ & $3.0{\cdot}10^{-5}$ & $5.9{\cdot}10^{-4}$ &$1.9{\cdot}10^{-3}$ & $3.4{\cdot}10^{-3}$ & $1.4{\cdot}10^{-4}$& $1.4{\cdot}10^{-3}$  & $2.6{\cdot}10^{-3}$ & $1.3{\cdot}10^{-2}$\\
\begin{math}{\sigma_s}[\frac{S}{mm}] \end{math} & $7.0{\cdot}10^{-5}$& $8.0{\cdot}10^{-5}$ & $9.0{\cdot}10^{-5}$ & $1.2{\cdot}10^{-4}$ & $1.6{\cdot}10^{-4}$  & $2.8{\cdot}10^{-4}$ & $3.1{\cdot}10^{-4}$  & $3.3{\cdot}10^{-4}$  \\
[.05ex] 
\hline 
\end{tabular}
\label{T} 
\end{table}

\begin{figure}[h]
  \centering
  \includegraphics[width=5.0in]{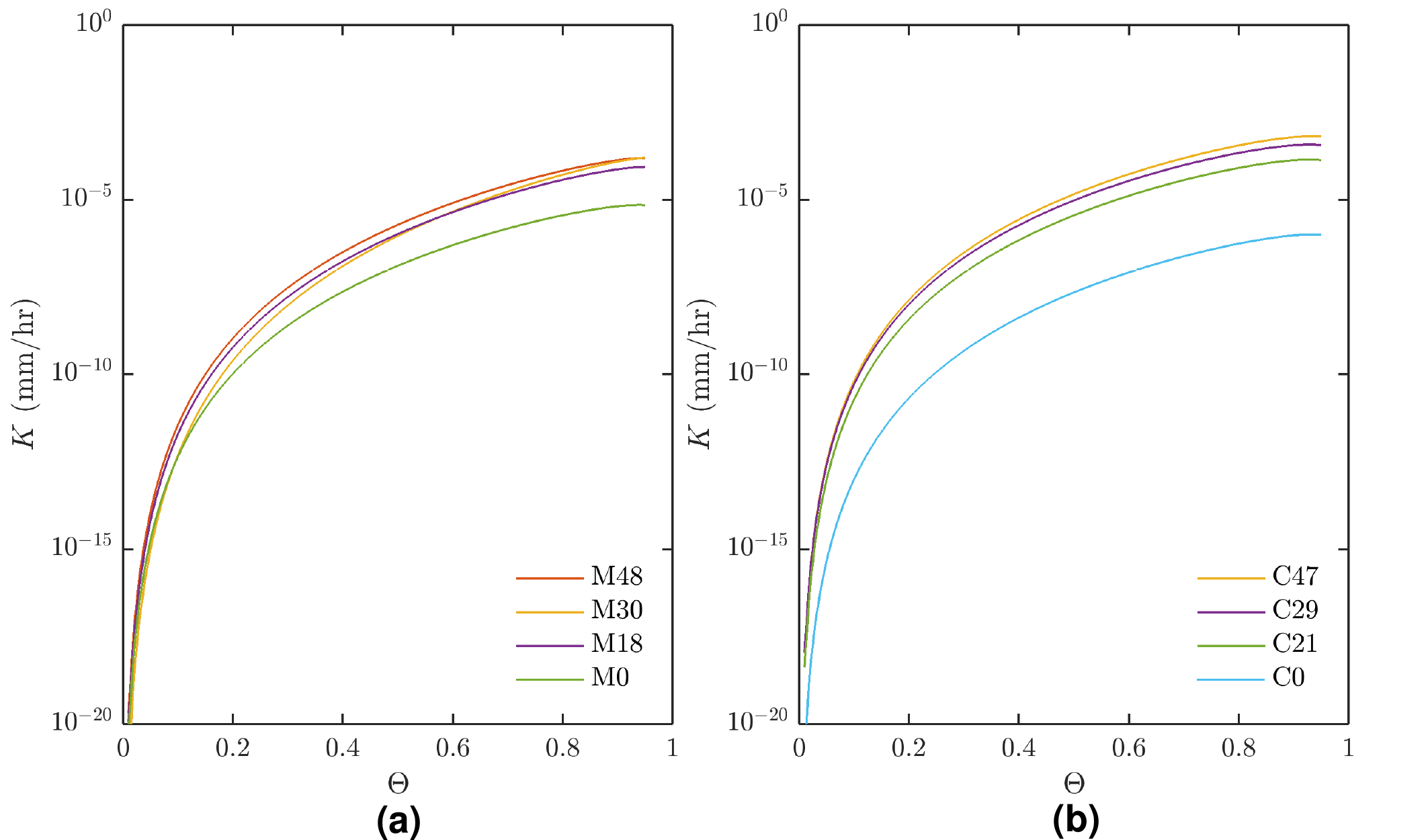}\\
  \caption{Hydraulic conductivity $K$ estimated using $S$ and moisture retention curves for undamaged and freeze-thaw damaged (a) mortar and (b) concrete.}\label{HC}
\end{figure}

We begin by determining $D$ using the power equation and assuming $\beta = 6.0$ \citep{leech2003unsaturated}.
Since $D$ is fully defined, we may use the hydraulic parameters ($S$, $n$, $\alpha$, and $\theta_s$) in Table \ref{T} to compute $K$ with Equation \ref{DKFP}.
The predictions of hydraulic conductivity for undamaged and freeze-thaw damaged mortar and concrete are shown in Figure \ref{HC}a and \ref{HC}b, respectively. 

Hydraulic conductivity predictions shown in Figure \ref{HC} exhibit similar trends as those reported in literature (cf.  \cite{Poyet2011}).
The range of $K$ considered here is $0 \leq \Theta \leq 0.97$, since no data was collected in obtaining the desorption isotherms in \citep{Smyl} between $0.97 < \Theta \leq 1.0$.
%As a result, $K$ predictions within 3\% of saturation diverge when this data is missing.
Nonetheless, $K$ values at 97\% saturation, for all materials considered,  are within one order of magnitude of $K_s$ values reported in Table \ref{T}, which is expected.
It is important to remark, however, that within 3\% of full saturation, the constraint $K(\Theta) \geq K(\Theta_{\delta^-})$ would need to be employed due to the tendency of the function toward zero.
%The scarcity of experimentally-obtained data for $K$ in cement based materials again prevents direct collaboration of results in Figure \ref{HC}.
%However, since $K$ follows the expected trend, $\frac{\partial K}{\partial \Theta} > 0$ \citep{durner1994hydraulic}, and $K_s$ was closely predicted near saturation, the predictions of $K$ are physically realistic.

\begin{remark}
{We would like to mention that the scarcity of data in the literature providing all variables required for relating unsaturated transport properties (for a given material) does add uncertainty to the generality of conclusions made herein.
While data obtained from freeze-thaw damaged material was used in this section, a large suite of data from  materials considering the effects of, for example, different mix designs, w/c ratios, admixtures, cements, curing conditions, etc. would be of significant research interest and increase the degree to which the results are verified.
}
\end{remark}

%%%%%%%%%%SECTION%%%%%%%%%%
\section{Model for {electrical conductivity} incorporating {pore-size distribution}}
\label{EC}

The model for $\sigma_b$ used here is similar to the model proposed by Mualem (Equation \ref{MualemSigma0}) using Brooks and Corey's expression for moisture retention.
Our model for $\sigma_b$ is written as

\begin{equation}
\sigma_b =\sigma_p  \theta^{(I +1)} F(m,q) 
\label{VGSigma}
\end{equation}

\noindent where $F(m,q)$ is written in terms of $\Omega= \Omega(m,q)$.
The exponent $(I+1)$ includes the addition of 1 due to the dependency of $\sigma_b$ on $\theta$ \citep{amente2000estimation}.
Moreover, Equation \ref{VGSigma} is expressed in terms of $\theta$, accounting for the material's open porosity \citep{HallHoff}.
{The saturation function} $F(m,q)$ was proposed in \citep{heimovaara1995assessing} using a modified van Genuchten approach {and is given by}

\begin{equation}
F(m,q) = \frac{(1-(1-\Theta^{1/m})^m)^2}{1-(1-\Theta^{1/q})^q}
\label{VFf}
\end{equation}

\noindent where $q = 1- \frac{2}{p}$ and $p >2.0$ are similar to the fitting parameters in the classic van Genuchten model.
We note that $F(m,q)$ requires that two moisture retention curves are fit for each estimation of $\sigma_b$.
%$F(m,q)$, is plotted in Figure \ref{f} for various levels of saturation, with material parameters $\theta_s = 0.15$ and $\alpha = 0.05$.
%
%\begin{figure}[H]
%  \centering
%  \includegraphics[width=3.5in]{fig4}\\
%  \caption{The saturation function, $F(m,q)$ plotted for various degrees of saturation, $\Theta$. }\label{f}
%\end{figure}
%
%Figure \ref{f} shows that $F(m,q)$ is strongly dependent on saturation and pore size distribution.
%High values of $m$, indicating a narrow $\Omega$, are shown to increase $F(m,q)$ faster than material with lower $m$.
%However, this does not necessarily mean that $\sigma_b$ is has a larger magnitude in materials with narrower $\Omega$ since in the present model is a function of several other variables: $\sigma_b = \sigma_b(\Omega, \sigma_p, \Theta, I)$.
Remaining to be determined in Equation \ref{VGSigma} is the pore solution conductivity $\sigma_p$, which is discussed in the following section.

%%%%SECTION%%%%
\subsection{Determining $\sigma_p$ with experimental data}
\label{sigmap}
There are multiple models for computing $\sigma_p$, (i) estimating the $\sigma_p$ from the concentration of ionic species, (ii) assuming $\sigma_p = \sigma_s$, and (iii)  using an empirical function to to estimate $ \sigma_p = \sigma_p(\Theta)$.
We first consider method (i).
Models estimating $\sigma_p$ as a function of OH$^-$, K$^+$, and Na$^+$ species were proposed in the literature (cf. \citep{weiss2012using,SNYDER}).
The models argued that under drying conditions, water is removed from the pore solution, resulting in stronger ionic concentration and higher $\sigma_p$.
However, precipitation of these ions during drying may result in poor prediction of $\sigma_p$ at low degrees of saturation \citep{rajabipour2007electrical}.
These models also require calibration factors for each species which may result in model error, especially at low degrees of saturation.
Due to the complexity of this model, we do not further consider it herein; we therefore move to analyzing points (ii) and (iii).

While data of $\sigma_p(\Theta)$ in cement-based materials is scarce, the non-linear function $\sigma_p(\Theta)$ has been established for several geologic materials \citep{schon2015physical}.
Due to the challenging experimental requirements of extracting $\sigma_p(\Theta)$, often $\sigma_p = \sigma_s$ is often taken for convenience \citep{amente2000estimation}.
It should be noted that estimating $\sigma_p=\sigma_s$ may lead to underestimation of $\sigma_p$ at low saturation levels since the electrical conductivity of the pore solution is at a minimum when the material is saturated \citep{weiss2012using}.
Therefore, this section considers $\sigma_p=\sigma_p(\Theta)$ using data from  \citep{rajabipour2007electrical}.

In developing a model for $\sigma_p(\Theta)$, we found that a simple exponential equation provided a very good fit to the experimental data.
The estimation is written as: $\sigma_p =\sigma_s\Theta^w$, where $w$ is a fitting parameter.
In Rajabipour and Weiss's study, the authors considered two cement pastes (w/c = 0.35 and 0.50); for both materials, $w = -0.95$ provided a good fit using a least-squares method.
The experimental data and fit for $\sigma_p$ is shown in Figure \ref{sigmap}.

\begin{figure}[h]
  \centering
  \includegraphics[width=3.25in]{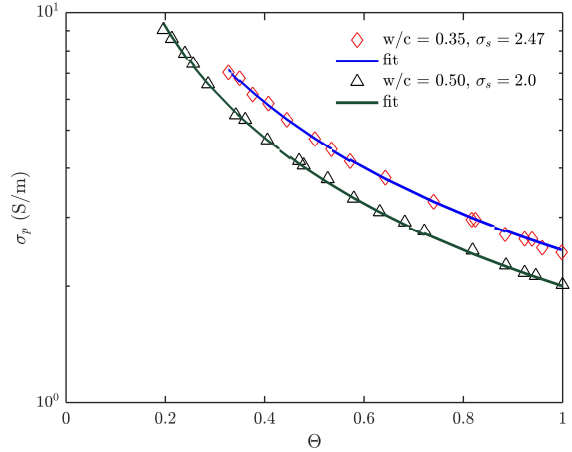}\\
  \caption{Experimental data of $\sigma_p$ from \cite{rajabipour2007electrical} and fitted curves as a function of $\Theta$. }\label{sigmap}
\end{figure}

%%%%%SECTION%%%%%%%%
\subsection{Application}

\begin{figure}[H]
  \centering
  \includegraphics[width=3.25in]{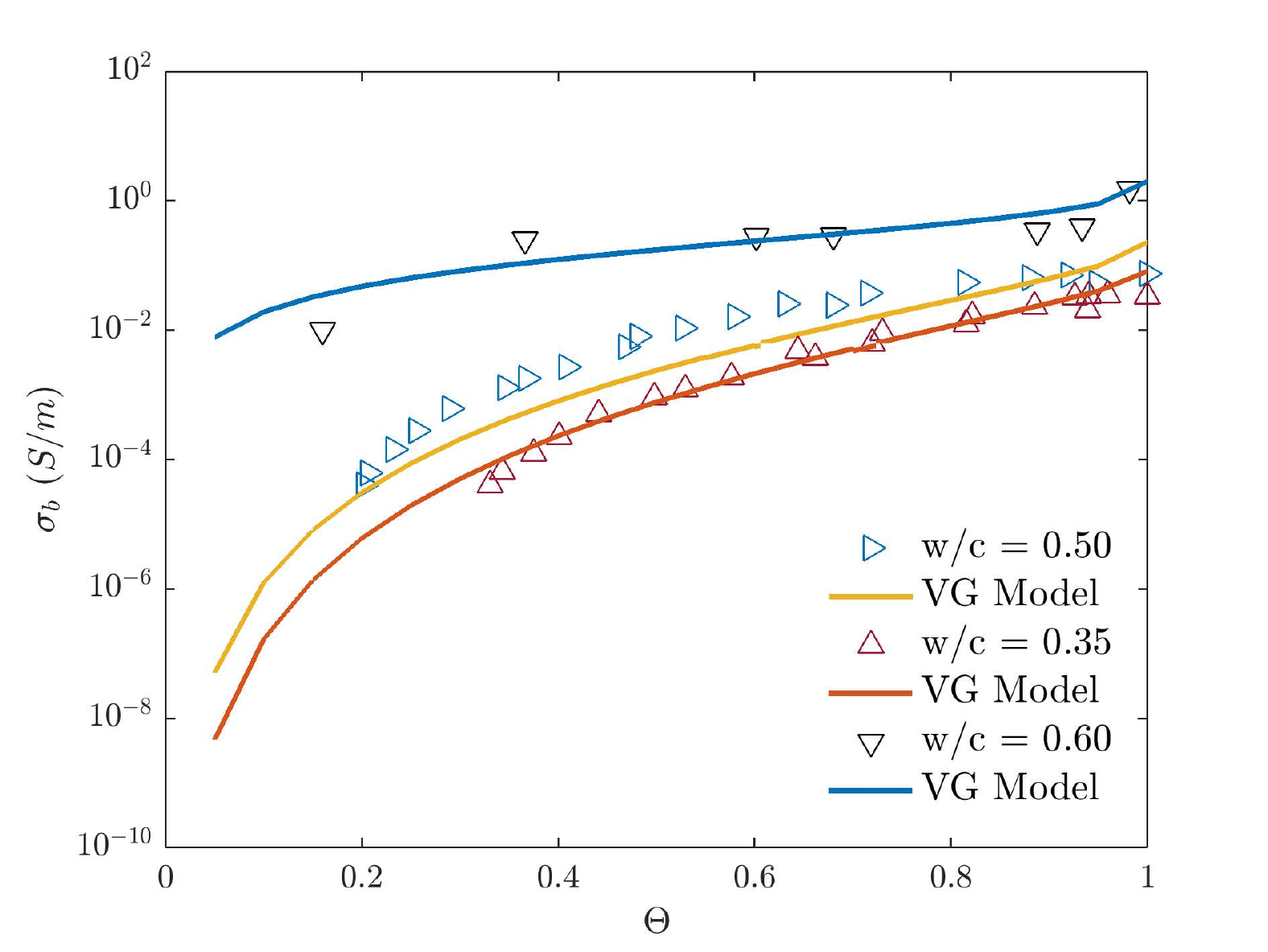}\\
  \caption{Experimental data of $\sigma_b$ from \citep{rajabipour2007electrical, Smyl3} plotted against $\Theta$. }\label{sigma}
\end{figure}

In this section, we study the feasibility of the van Genuchten-Mualem model for $\sigma_b$ by comparing it to experimentally-measured $\sigma_b$ of cement paste.
Three OPC cement pastes were selected in the analysis with w/c = 0.60, 0.50, and 0.35 (with 5\% silica fume).
The data for w/c = 0.60 was obtained from \cite{Smyl3} and the data for w/c = 0.50 and 0.35 was obtained from \cite{rajabipour2007electrical}.
While $\sigma_p(\Theta)$ was known for the pastes with w/c = 0.50 and 0.35, it was not known for w/c = 0.60.
We therefore approximate $\sigma_p(\Theta)$ for w/c = 0.60 using the same curve for w/c = 0.50.
The unknown parameter $I$ was selected using a least-square fit to the experimental data.
The model for $\sigma_b$ for the three cement pastes is plotted in Figure \ref{sigma}.

Figure \ref{sigma} demonstrates that the model used to estimate $\sigma_b$ compares well with the experimental data.
We observe that the model predicting $\sigma_b$ for the paste with w/c = 0.50 underestimates $\sigma_b$  at intermediate degrees of saturation and near full saturation.
This is an artifact attributed to the lack of data in the desorption isotherm near saturation and at $\Theta < 0.50$ resulting in a poor fit of the moisture retention curve.
Even so, the model was within an order of magnitude of the experimental data, supporting the feasibility of the model for cement-based materials.

%%%%%SECTION%%%%%%%%
\section{Relating {hydraulic and electrical conductivity}}
\label{AL}
We have shown that $K$ and $\sigma_b$ may be determined without directly measuring either quantity, which is convenient since measurement of $K$ experimentally is generally not feasible for cement-based materials.
With these parameters defined, we now aim to link $K$ and $\sigma_b$.
One advantage of using the van Genuchten-Mualem-based models to predict $K$ and $\sigma_b$ is that both models use the same variables describing $\Omega$ and $I$.
That is, for a given material, these values are equivalent.
Starting from Equation \ref{MualemSigma} we may solve for $I$ explicitly

\begin{equation}
I= \frac{\ln\Big(\frac{K}{K_s((1 - \Theta^{1/m})^m - 1)^2}\Big)}{\ln(\Theta)}.
\label{Isolve}
\end{equation}

Similarly, we may solve Equation \ref{VGSigma} for $I$, which is written 

\begin{equation}
I=\frac{ \ln \Big(\frac{\sigma_b((1 -\Theta^{1/q})^q - 1)}{\sigma_p(1 - \Theta^{1/m})^m - 1) }\Big)}{\ln(\theta)} - 1.
\label{Isolve2}
\end{equation}

Since $I$ is equivalent in both models \citep{mualem1991theoretical}, we may equate both terms on the right-hand side of Equations \ref{Isolve} and \ref{Isolve2}

\begin{equation}
\frac{\ln\Big(\frac{K}{K_s((1 - \Theta^{1/m})^m - 1)^2}\Big)}{\ln(\Theta)}=\frac{ \ln \Big(\frac{\sigma_b((1 -\Theta^{1/q})^q - 1)}{\sigma_p(1 - \Theta^{1/m})^m - 1) }\Big)}{\ln(\theta)} - 1.
\label{Isolve3}
\end{equation}

The above expression may be solved in terms of either $\sigma_b$ or $K$, providing expressions relating the two material properties.
In others words, we may write $K = K(\sigma_b)$ or $\sigma_b = \sigma_b(K)$.
By solving for $\sigma_b$ and $K$, we obtain

\begin{equation}
K = K_s\exp\Bigg(\ln(\Theta)\frac{\ln\Big(\frac{\sigma_b}{\sigma_p}\frac{(1 - \Theta^{1/q})^q - 1}{(1 -\Theta^{1/m})^m-1 }\Big)}{\ln(\theta) }  - 1 \Bigg)\cdot((1 - \Theta^{1/m})^m - 1)^2
\label{Isolve4}
\end{equation}

and

\begin{equation}
\sigma_b = \sigma_p\exp \bigg(\ln(\theta)\frac{\ln \Big (\frac{K}{K_s}((1 - \Theta^{1/m})^m - 1)^2\Big)}{ \ln(\Theta) + 1}\Bigg)\cdot \frac{(1 - \Theta^{1/m})^m - 1}{(1 - \Theta^{1/q})^q - 1}
\label{Isolve5}
\end{equation}

However, since $K_r = K/K_s$, we may also write

\begin{equation}
K_r = \exp\Bigg(\ln(\Theta)\frac{\ln\Big(\frac{\sigma_b}{\sigma_p}\frac{(1 - \Theta^{1/q})^q - 1}{(1 -\Theta^{1/m})^m -1}\Big)}{\ln(\theta) }  - 1 \Bigg)\cdot((1 - \Theta^{1/m})^m - 1)^2
\label{Isolve6}
\end{equation}

The expressions for $K$, $K_r$, and $\sigma_b$ in Equations \ref{Isolve4}-\ref{Isolve6} are independent of the parameter $I$.
This is convenient, since $I$ cannot be determined unless either $K$ and $\sigma_b$, as well as $\Omega$, are explicitly known.
$D$ may also be determined from these expressions (i.e by substituting the expression $D= Kg$, where $g$ is given in Section \ref{DFKG}).
As in Equations \ref{DKF} and \ref{DKFP}, Equations \ref{Isolve4}-\ref{Isolve6} tend towards zero near saturation, and should therefore employ the constraints $\sigma_b(\Theta) \geq \sigma_b(\Theta_{\delta^-})$, $K=(\Theta) \geq K(\Theta_{\delta^-})$, and $K_r(\Theta) \geq K_r(\Theta_{\delta^-})$.

\begin{remark}
{It should be recapitulated that $K$ and $\sigma_b$ have phenomenological differences.
Perhaps the primary difference is that $K$ has dependence on capillary forces in unsaturated pores whereas $\sigma_b$ has dependence on electrical conductivity of the pore solution \citep{Scherer,ghasemzadeh2016comparison}.
The simplistic nature of Equations \ref{Isolve4}-\ref{Isolve6}, however, places restrictions on the ability to physically interpret the role ions in pore solution have on $K(\sigma_b)$ and $\sigma_b(K)$.
In a sensitivity analysis, using available data from \cite{rajabipour2007electrical}, it was found that Equations \ref{Isolve4}-\ref{Isolve6} had weak dependence on $\sigma_p$.
This indicates that, \emph{for these relations}, $K(\sigma_b)$ and $\sigma_b(K)$ are strongly dependent on pore connectivity and $\Omega$ and secondarily dependent on $\sigma_p$.
However, in cases where additional accuracy in modeling $K(\sigma_b)$ and $\sigma_b(K)$ is required, a more advanced model including the chemistry inherent in $\sigma_p$ may be needed.}
\end{remark}

%%%%%%%%%%%SUBSECTION%%%%%%%%%%%%%%
\subsection{Application: $K_r = K_r(\sigma_b)$}

\begin{figure}[h]
  \centering
  \includegraphics[width=3.25in]{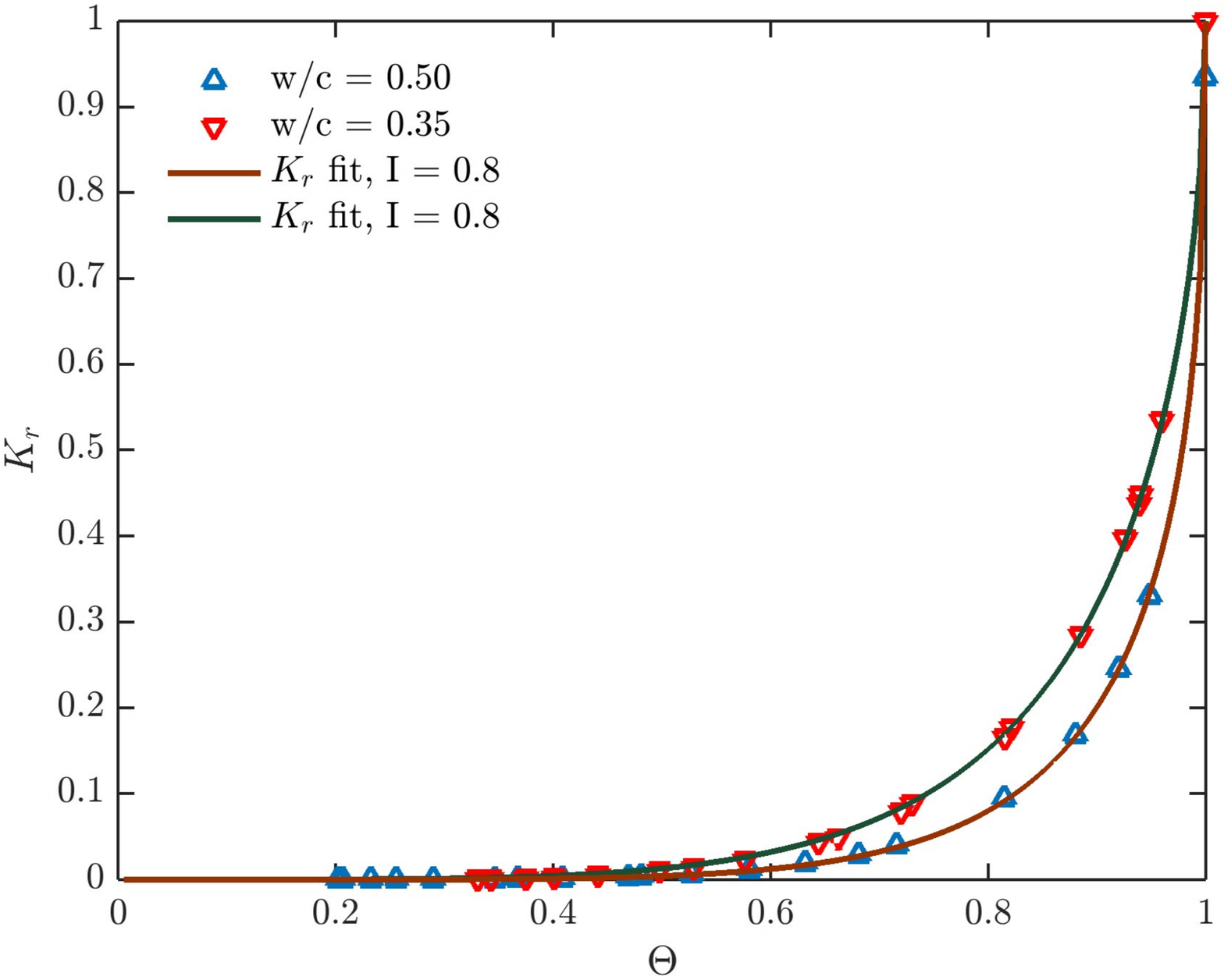}\\
  \caption{Prediction of $K_r(\Theta)$  using data from \citep{rajabipour2007electrical}. }\label{prediction1}
\end{figure}

In this application, we again use the experimental data from \cite{rajabipour2007electrical} and the fitted equation for $\sigma_p$ provided in Section \ref{EC} to predict $K_r = K_r(\sigma_b)$ (Equation \ref{Isolve6}).
We predict $K_r$ rather than $K$, since $K_s$ data for these cement pastes was not known.
{To illuminate the functional relation of $K_r$ on $\Theta$, we first compute $K_r(\Theta)$ using the basic form of the van Genuchten Equation (Equation \ref{MualemSigma}).}
Figure \ref{prediction1} shows the model's predictions and fitting of $K_r$ using the van Genuchten-Mualem model and using $I=0.8$ which resulted in a good fit.

\begin{figure}[h]
  \centering
  \includegraphics[width=3.25in]{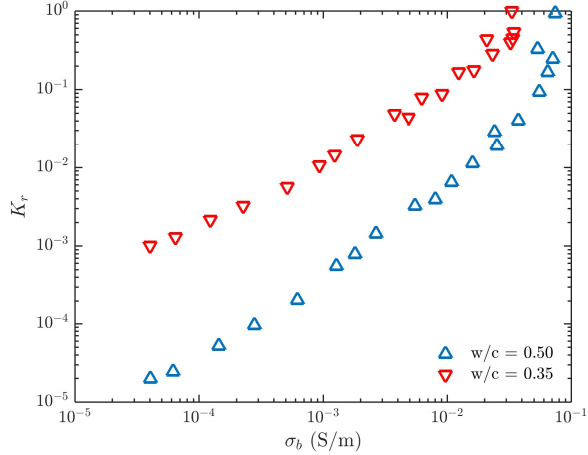}\\
  \caption{Prediction model of $K_r$ as a function of electrical measurements ($\sigma_b$). }\label{SKr}
\end{figure}

We remark that, while it is certainly feasible to use Equation \ref{Isolve2} to determine $I$ from electrical data, it was simpler to approximate it by fitting as shown in Figure \ref{prediction1}.
%Indeed, the curves using $I$ as a fitting parameter (van Genuchten-Mualem model) match the prediction of $K_r$ using electrical measurements quite well.
We now consider the estimation of $K_r(\sigma_b)$, which doesn't require computing $I$.
The $K_r(\sigma_b)$ prediction model (Equation \ref{Isolve6}) is plotted in Figure \ref{SKr} against the experimental electrical measurements showing a  log-log linear trend for intermediate levels of $\sigma_b$.

We would like to mention again that no data is available for cement-based materials in the literature to corroborate Figures \ref{prediction1} and \ref{SKr}.
However, the nonlinear trend observed for intermediate degrees of saturation should be expected.
This trend results from the well-known exponential dependence of both $\sigma_b$ and $K_r$ on {saturation in porous materials}.
{For example,} the log-log linear trend between $K$ and resistivity ($\rho_b = 1/\sigma_b$)  has been observed in geologic materials such as aquifers \citep{khalil2009,mazavc1985,kelly1984}, sandstone \citep{revil2014electrical}, and fractured rocks \citep{frohlich1996}.
{It is also interesting to note that rate of change in Figure \ref{SKr} appears to be proportional to the w/c ratio, i.e $\frac{\partial K_r}{\partial \sigma_b} \propto w/c$.
%Such a result may indicate that the dependence of $K_r$ on $\sigma_b$ increases with the w/c ratio.
However, such an interpretation should be taken with caution since only two materials are studied.}
%Moreover, such an interpretation should not be extended to the dependence of $K$ on $\sigma_b$ since the relation $K = K_sK_r$ does not guarantee the former.

%%%%%%%%%%%SUBSECTION%%%%%%%%%%%%%%
\subsection{Application: $\sigma_b = \sigma_b(K)$}
In this section we predict $\sigma_b$ from {$K$ using data from undamaged and damaged material provided in Section \ref{Dam1}}.
Since $K$ for cement-based materials is generally not available experimentally \citep{zhou2015unified}, predicting $\sigma_b$ from $K$ requires indirect estimation of $K$.
In this application we use information from $S$ and $\Omega$ to estimate $K$, which follows the path detailed in Section \ref{DFKG}: ($S \to D , (D,\Omega) \to K)$.

Using $S$ to determine $D$ is a rather straight forward step in this process.
However, the application of Equation \ref{Isolve5} is not as straight forward as using Equation \ref{Isolve6} to predict $K_r = K_r(\sigma_b)$ -- this will be explained in the following.
When testing Equation \ref{Isolve5} in preliminary analysis, it was found the dependence on $\ln(\theta)$ resulted in unrealistic prediction of $\sigma_b$ near saturation (i.e $\sigma_b(\theta= \theta_s) >> \sigma_s$).
It was found that, by substituting $\ln(\Theta)$ for $\ln(\theta)$ in Equation \ref{Isolve5}, the solutions were physically realistic.
Analytically, $\ln(\Theta \Rightarrow 1.0) \Rightarrow 0$ whereas $\ln(\theta  \Rightarrow \theta_s) \Rightarrow$ negative value; the resulting negative solutions from using $\ln(\theta)$ in the exponent of Equation \ref{Isolve5} led to divergence of $\sigma_b$ near saturation.
Substituting $\ln(\Theta) $ resulted in a prediction of $\sigma_b=\sigma_s$ on the order of magnitude of the experimentally-measured $\sigma_s$ near saturation and more realistic across the range of $\Theta$, this substitution is used herein.
%\footnote{\textcolor{blue}{The modified expression for $\sigma_b$ is written as: $\sigma_b = \sigma_p\exp \bigg(\ln(\Theta)\frac{\ln \Big ({K}/{K_s}((1 - \Theta^{1/m})^m - 1)^2\Big)}{ \ln(\Theta) + 1}\Bigg)\cdot \frac{(1 - \Theta^{1/m})^m - 1}{(1 - \Theta^{1/q})^q - 1}$.}}.

\begin{figure}[h]
  \centering
  \includegraphics[width=5.25in]{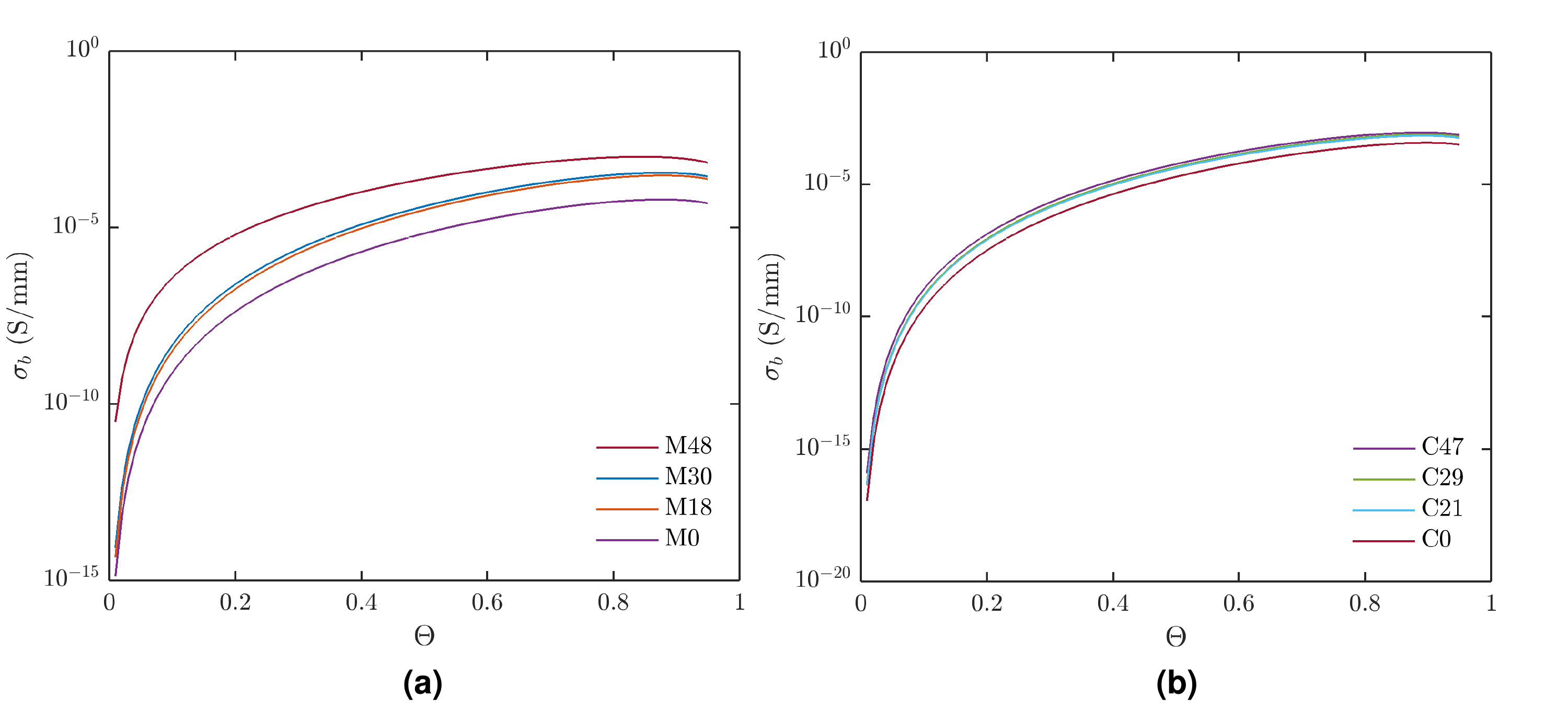}\\
  \caption{Prediction of $\sigma_b$ from $K$ determined from $S$ and $\Omega$ for undamaged and damaged (a) Mortar and (b) Concrete. Results are plotted against $\Theta$.}\label{S1}
\end{figure}

Implementing Equation \ref{Isolve5} requires knowledge of $\sigma_p$, due to the lack of data for $\sigma_p(\Theta)$, we used the simplification $\sigma_p = \sigma_s$.
We would like to mention that the preliminary analysis showed weak dependence of $\sigma_b$ on $\sigma_p(\Theta)$ using the function presented in Section \ref{sigmap}.
This indicated the the approximation of $\sigma_p = \sigma_s$ was reasonable.
Implementing Equation \ref{Isolve5} also requires knowledge of $K_r=K/K_s$, where $K$ is known from the procedure highlighted above.
Rather than using experimental values for $K_s$, we used the relation $K_s = \mathrm{max}(K)$.
%This conveniently eliminates the need to directly measure $K_s$ using this method.
The predictions of $\sigma_b(K)$ for the materials in Section \ref{Dam1} are shown in Figure \ref{S1}, plotted first as a function of $\Theta$ for $\Theta < 0.97$ (again due to lack of data near saturation).

While there is no experimental data to corroborate $\sigma_b(K)$ in Figure \ref{S1}, the values of $\sigma_b$ near saturation are within one order of magnitude of the experimentally-measured values reported in Table \ref{T}.
Moreover, the curves generally follow the expected trends in the intermediate range of $\Theta$ (cf. cement paste results reviewed in \citep{li2016effect}).
The functional tendency of Equation \ref{Isolve5} to towards zero near $\Theta = 0.97$ is also observed in Figure \ref{S1}, since no constraint (i.e. $\sigma_b(\Theta) \geq \sigma_b(\Theta_{\delta^-})$) was applied.

\begin{figure}[h]
  \centering
  \includegraphics[width=5.25in]{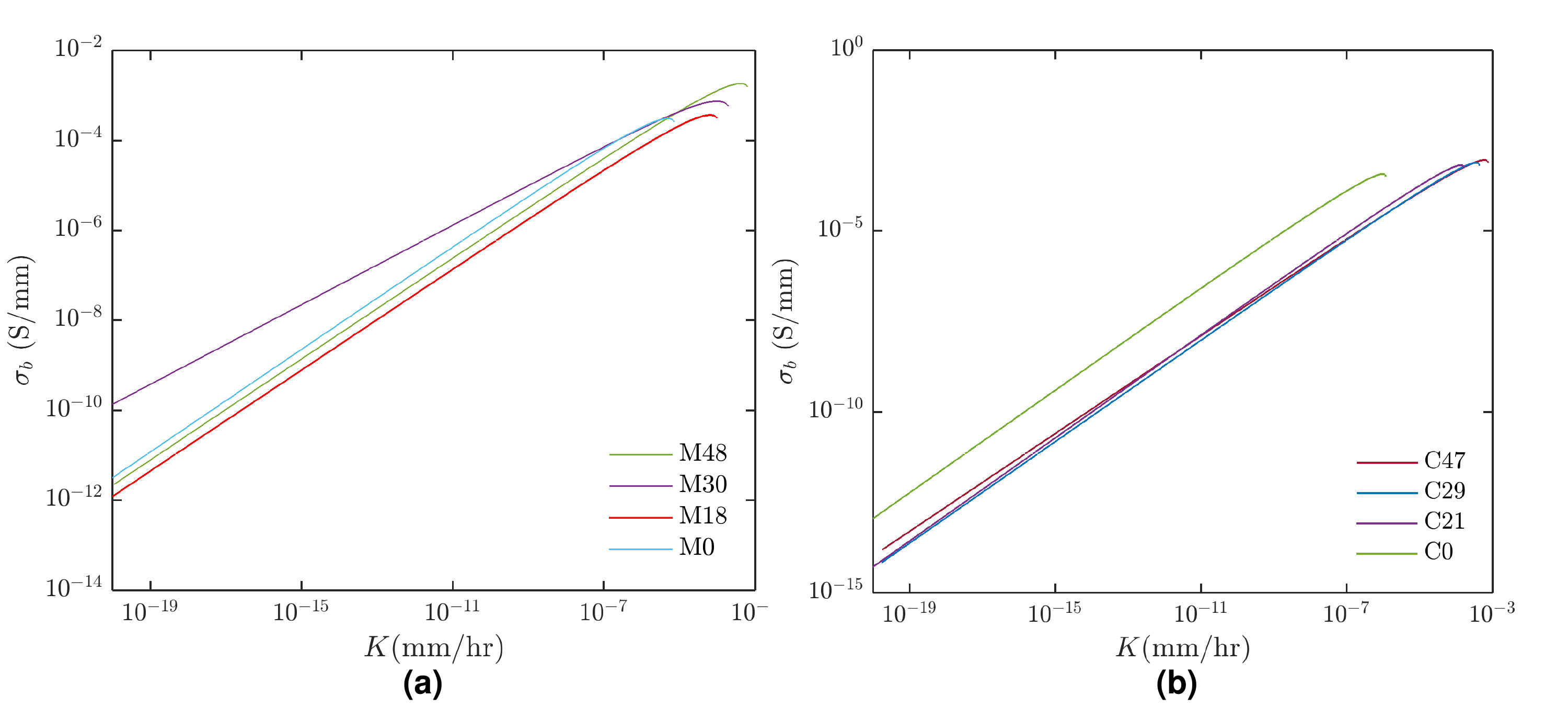}\\
  \caption{Prediction of $\sigma_b$ from $K$ determined from $S$ and $\Omega$ for undamaged and damaged (a) Mortar and (b) Concrete.  Results are plotted against $K$.}\label{S2}
\end{figure}

While the dependence of $\sigma_b$ on $\Theta$ for materials with distributed damage is not well established, it is expected that higher damage levels will increase $\sigma_b(\Theta)$.
This is expected since the the overall porosity is higher and the pores are more well-connected \citep{smyl2017D}.
Indeed, it is clear in Figure \ref{S1}a that $\sigma_b(\Theta)$ increases proportional to damage.
This is also seen in Figure \ref{S1}b, albeit more subtly due to the larger range of $\sigma_b$.
%We would like to mention here that the behavior of $\sigma_b$ near saturation decreases slightly, which is due to the lack of moisture retention data within 3\% of saturation and using $\beta = 6.0$ in determining $D$ (higher $\beta$ increases $K$ and $\sigma_b$ near saturation).
%Therefore only $\Theta < 0.97$ is shown in Figure \ref{S1}.
Next, we investigate the dependence of $\sigma_b$ on $K$, which is shown in Figure \ref{S2}.

The log-log linear trend relating $\sigma_b$ and $K$, as shown in Figure \ref{SKr}, is again observed.
%We remark that this trend was also noted in studies of geologic materials, such as fractured granite \citep{frohlich1996} or single apertures \citep{brown1989transport}.
The curves shown in Figure \ref{S2} result from the highly non-linear dependence of $\sigma_b$ on $\Theta$, $\Omega$, $\theta_s$, $\theta_i$, $K_s$, $\sigma_s$, and $S$.
In particular, the van Genuchten-Mualem parameters $m$ and $\alpha$ have a significant effect on the slope of the plots, while $S$, $\theta_s - \theta_i$, $K_s$, and $\sigma_s$ control the vertical position of the curves.
In other words, $\Omega$ controls the rate dependence of $\sigma_b$ on $K$ (and vice-versa).
These aspects will be further detailed using a simplified model in the following section.

%%%%%%%%%%%%SECTION%%%%%%%%%%%%%%%
\subsection{Simplified models for $K(\sigma_b)$ and $\sigma_b(K)$}
The log-log linear trends shown in Figures \ref{SKr} and \ref{S2} for $K=K(\sigma_b)$ and $\sigma_b = \sigma_b(K)$, respectively, resulted from the dependencies on $\Omega$, $S$, $\theta_s$, $\theta_i$, $\Theta$, $K_s$, and $\sigma_s$.
Quantifying the roles of each parameter in the analytical models (Equations \ref{Isolve4}-\ref{Isolve6}) is cumbersome due to the complex functional relations of the numerous variables.
Moreover, the use of constraints on $K$ and $\sigma_b$ are required since the analytical models tend to zero close to $\Theta =1.0$.

In this section we present a simplified model to describe $K=K(\sigma_b)$  and $\sigma_b = \sigma_b(K)$.
The simplified model is designed to reduce the number of variables, yet closely approximate the curve shape of the analytical models in Section \ref{AL}.
This proposed model is written as 

\begin{equation}
\sigma_b = \kappa K \Theta^\gamma 
\label{Simple}
\end{equation}

or in terms of $K$

\begin{equation}
K = \frac{\sigma_b}{\kappa \Theta^\gamma}
\label{Simple2}
\end{equation}

\noindent where  and $\gamma(-)$ is a fitting parameter and the constant $\kappa~\mathrm{(\frac{S\cdot h}{mm^2})}$ may be determined by analyzing the saturated form of Equation \ref{Simple}:

\begin{equation}
\sigma_b(\Theta = 1.0) = \sigma_s= \kappa  K_s
\label{Simpled}
\end{equation}

which leads to

\begin{equation}
\kappa = \frac{\sigma_s}{K_s}.
\label{Simpled2}
\end{equation}

Equation \ref{Simple} and the analytical model from Section \ref{AL} (Equation \ref{Isolve5}) are monomials, resulting in log-log linear trends for $0 <\Theta<1.0$.
In Equations \ref{Simple} and \ref{Simple2}, $\gamma$ controls the slope and $\kappa$ shifts the curve vertically.
We remark that for $\sigma_b$ and $K$ to be physically realistic $\sigma_b$ and $K$ must increase with saturation (i.e $\frac{\partial \sigma_b}{\partial \Theta}~ \mathrm{and} ~\frac{\partial K}{\partial \Theta} >0$).
This also implies that $\frac{\partial \sigma_b}{\partial K} ~\mathrm {and}~ \frac{\partial K}{\partial \sigma_b}> 0$.
Further, we may solve for the unknown $\gamma$ by writing

\begin{equation}
\gamma = \frac{\ln \Big( \frac{ K_s \sigma_b }{K \sigma_s}\Big)}{\ln(\Theta)}= \frac{\ln \Big( \frac{ \sigma_r }{K_r}\Big)}{\ln(\Theta)},~\sigma_r=\frac{\sigma_b}{\sigma_s}.
\label{ggamma}
\end{equation}

From Equation \ref{ggamma}, we observe that the sign of $\gamma$ is determined\footnote{$\gamma$ may be written in terms of the van Geunchten parameters using Eq. \ref{ggamma} and substituting $K_r$ (Eq. \ref{MualemSigma}) and $\sigma_r=\frac{\sigma_b}{\sigma_s}$  yielding: $\gamma =\ln((1 - (1-\Theta^{1/p})^p)^{-1})\ln(\Theta)^{-1}$  with the simplifying assumptions $\sigma_b \propto \Theta^I$ and $\sigma_p = \sigma_s$.} by  $ \frac{\sigma_r}{K_r} > 0, \neq 1.0$.
For a fixed value of $\frac{\sigma_r}{K_r}$, $\gamma$ is negative when $\frac{\sigma_r}{K_r} > 1.0$, and is otherwise positive.
In this case, $\gamma$ is dependent on $\Theta$ when $\frac{\sigma_r}{K_r}$ is constant and $\gamma$ increases (or decreases, depending on the sign) exponentially near saturation.
However, a fixed value of $\frac{\sigma_r}{K_r}$ is a unique instance and is not physically realistic, since $\gamma$ must be constant in order for the condition that $K$ and $\sigma_b$ are log-log linear to hold.
It is apparent that $\frac{\sigma_r}{K_r}$ varies with $\Theta$ by writing Equation \ref{ggamma} in a simpler form: 

\begin{equation}
\frac{\sigma_r}{K_r}= \Theta^\gamma.
\label{gammar}
\end{equation}

In the physically-realistic case that $\frac{\sigma_r}{K_r}$ changes with $\Theta$, $\gamma$ is constant and may take a positive or negative sign.
When $\gamma$ is positive, $ K_r > \sigma_r$ and vice versa.
Since $\gamma\neq \gamma(\Theta)$, we surmise that $\gamma$ does not include the effects of tortuosity ($\tau$) and pore connectivity ($\zeta$), as explained in the following.

It is well known that $\tau(\Theta)$ increases and $\zeta(\Theta)$ decreases with decreasing $\Theta$ \citep{smyl2017D,AkhavanCCR,HallHoff}, yet it was shown that $\gamma$ is invariant  of $\Theta$.
This would indicate that information of $\tau$ and $\zeta$  are stored solely within $K$ and $\sigma_b$ (or $K_r$ and $\sigma_r$).
In this case, it is implied in Equation \ref{gammar} that the effects of $\tau$ and $\zeta$ are canceled in dividing $\sigma_r$ by $K_r$, yielding $\frac{\sigma_r}{K_r}$ dependence on the degree of saturation and $\gamma$ only.

Therefore, the physical interpretations of $\gamma$ are either (i) $\gamma$ is only dependent on $\Omega$ or (ii) $\gamma$ is an interpretation of $\Omega$.
Consequently, the lack of dependence of $\gamma$ on $\Theta$ is somewhat intuitive since $\Omega$ is also invariant of $\Theta$ (when moisture hysteresis, chemical, and temperature effects on pore-size distribution are neglected).
Since $\tau$ and $\zeta$ are inferred to be already contained within $K$ and $\sigma_b$, we only need $\Omega$ to describe $\frac{\sigma_r}{K_r}$.
Further, when $\Omega$, $K_s$, and $\sigma_s$ are known,  $\frac{\sigma_b}{K}$ is defined.
We note that due to the simplifications of the model proposed in this section and the general lack of experimental data corroborating $K$ and $\sigma_b$, further research is required to confirm the roles of $\tau$ and $\zeta$ with respect to $\frac{\sigma_b}{K}$.

\begin{remark}
{The use of the proposed simplified model should be done with some caution when characterizing materials with various constitutions.
One inherent uncertainty/limitation of the simplified model is that it employs the assumption that the conductivity of the pore solution is constant ($\sigma_p = \sigma_s$).
While this may be a reasonable assumption for some materials, for others it may not be.
In such cases, where $\sigma_b(K)$ and $K(\sigma_b)$ may have sensitivity to $\sigma_p$ outside application-specific constraints, discretion should be used.}
\end{remark}

%%%%%%%%%%%%SECTION%%%%%%%%%%%%%%
\section{Conclusions}
In this work, we aimed to quantify the relation between unsaturated electrical ($\sigma_b$) and hydraulic ($K$) conductivity.
Relating these two quantities is of particular interest, because obtaining $K$ is often experimentally impractical while measuring $\sigma_b$ is rather straight forward.
We began by using the van Genuchten-Mualem model to determine the roles of pore-size distribution and saturation on $K$ and $\sigma_b$.
Experimental data was then used to determine $K$ and $\sigma_b$ indirectly.
The models were then confirmed to be reasonably well corroborated with the scarcely-available experimental data.
{It was noted that (a) experimentally-measured $K$ of cement-based materials is virtually non-existent; therefore, more data is required to determine the accuracy $K$ and $\sigma_b$ prediction models and (b) additional transport property data from other cement-based materials considering different mix design constituents would be of significant research interest and improve understanding of the efficacy of the models proposed herein.}

The models developed for $\sigma_b$ and $K$ were analytically linked such that $\sigma_b = \sigma_b(K)$ or $K = K(\sigma_b)$.
The analytical models showed that $\sigma_b(K)$ or $K(\sigma_b)$ followed a log-log linear trend.
Due to the complexity of these models -- resulting from their functional dependencies on pore-size distribution{and transport properties}  -- we proposed a simplified model that closely approximates the more complex analytical models.
Interpretations from the simplified model suggest that the ratio of relative electrical conductivity to relative hydraulic conductivity is only dependent on saturation and pore-size distribution.
Further research and experimental data is required to quantify the roles of tortuosity and pore connectivity in relating $K$ and $\sigma_b$.

%%%%%%%%%%%%SECTION%%%%%%%%%%%%%%
\section*{Acknowledgments}
This work was supported by the Department of Mechanical Engineering at Aalto University and a Fulbright Grant.
The author would also like to acknowledge Mohammad Pour-Ghaz at North Carolina State University for many useful discussions.

\bibliographystyle{chicago}
\bibliography{bibliography}

\end{document}